\documentstyle[twocolumn,seceq]{jpsj}
\def\geqap{\,\raise 2pt \hbox{$>\kern-11pt \lower 5pt \hbox{$\sim$}$}\,}
\def\leqap{\,\raise 2pt \hbox{$<\kern-10pt \lower 5pt \hbox{$\sim$}$}\,}
\def\runtitle{Low-Energy Charge Dynamics of NaV$_2$O$_5$}
\def\runauthor{Satoshi {\sc Nishimoto} and Yukinori {\sc Ohta}}
\title{A Model Study of the Low-Energy Charge Dynamics of NaV$_2$O$_5$}
\author{Satoshi {\sc Nishimoto}$^{1}$ and Yukinori {\sc Ohta}$^{1,2}$}
\inst
{$^1$ Graduate School of Science and Technology, Chiba University, 
Inage-ku, Chiba 263-8522\\
$^2$Department of Physics, Chiba University, Inage-ku, Chiba 263-8522}
\recdate{September 11, 1998}
\abst
{An exact-diagonalization technique on small clusters is used to 
calculate the dynamical density correlation functions of the dimerized 
$t$-$J$ chain and coupled anisotropic $t$-$J$ ladders (trellis lattice) 
at quarter filling, i.e., the systems regarded as a network of pairs 
(dimers or rungs) of sites coupled weakly via the hopping and exchange 
interactions.  We thereby demonstrate that the intersite Coulomb 
repulsions between the pairs induce a low-energy collective mode in the 
charge excitations of the systems where the internal charge degrees of 
freedom of the pairs play an essential role.  
Implications to the electronic states of NaV$_2$O$_5$, i.e., fluctuations 
of the valence state of V ions and phase transition as a charge ordering, 
are discussed.
}
\kword
{NaV$_2$O$_5$, charge dynamics, $t$-$J$ model, 
intersite Coulomb, coupled ladders, trellis lattice}
\begin{document}
\sloppy
\maketitle

A compound $\alpha'$-NaV$_2$O$_5$ discovered as an inorganic 
`spin-Peierls' system\cite{isobe} has attracted much attention 
because of its intriguing charge degrees of freedom related to 
fluctuations of the valence state of V ions.  It has been reported, 
e.g., that the ratio of the size of the spin gap at $T=0$ K to 
the transition temperature is largely deviated from the 
BCS mean-field ratio as $2\Delta/k_{\rm B}T_c=6.5$ with 
$\Delta=9.8$ meV\cite{fujii,yosihama} and $T_c=34$ K,\cite{isobe} 
that a large amount of entropy is released at the phase transition 
in contrast to the case in CuGeO$_3$,\cite{koppen,powell} and that 
anomalous behavior of the dielectric constant\cite{sekine,smirnov} 
along the {\mib a}-axis occurs at $T_c$ while no anomaly is observed 
along the {\mib b}-axis.\cite{sekine}  
Below $T_c$, a huge increase in the thermal conductivity has been 
reported to occur,\cite{vasilev} and with a direct evidence, the 
charge-ordering scenario for the mechanism of the phase transition 
has been put forward from NMR experiment.\cite{ohama}  Some possible 
ordering patterns of charges at $T<T_c$ have been suggested 
from theories.\cite{thalmeier,seo,nishimoto3,mostovoy}.  
It has also been pointed out that the $\sim$1 eV features of the 
observed optical conductivity\cite{opt1,opt2} cannot be interpreted 
without assuming the system to be charge disproportionated even 
above $T_c$,\cite{opt2,nishimoto4} and to be reconciled with the 
picture of the uniform valence of V ions which X-ray structural 
analyses have shown,\cite{xray1,xray2,xray3} we have argued that 
the rapid charge oscillation should exist in the system above 
$T_c$.\cite{nishimoto4}  

There are also a number of experiments detecting fluctuations above 
$T_c$.  A recent line-shape analysis in the X-ray diffuse-scattering 
experiment\cite{ravy} suggests the pretransitional structural 
fluctuations to appear, where the correlation lengths start to 
increase around $\sim$50 K, $\sim$70 K, and above 90 K for the 
${\mib c}$-, ${\mib a}$-, and ${\mib b}$-axes, respectively, and 
diverge as a power low by lowering temperature down to $T_c$.  
There has been an independent observation of the presence of critical 
scattering from soft phonons above $T_c$.\cite{nakao}  
Such a pretransitional effect has also been noticed in the ultrasonic 
experiment,\cite{fertey} where the anomalous behavior of sound 
velocity starts to develop at $\sim$80 K.  Anomalous increase in the 
oscillator strength of the low-energy continuum observed at 
$T\simeq 150$ K in the optical conductivity experiment\cite{opt2} 
might also be related to this effect.  
All of these experimental data seem to suggest that to clarify 
the mechanism of the phase transition of this compound it should be of 
primary importance to study the nature of the fluctuations and ordering 
of electronic charges; we therefore consider this issue in the present 
paper.  

The high-energy electronic states of the V-O network of the 
(V$_2$O$_3$)$^{3+}$ layer may be described by the $d$-$p$ model with 
strongly-correlated $d_{xy}$-orbitals of V ions bridged by the 
$p_x$- and $p_y$-orbitals of O ions.  Our analysis\cite{nishimoto3} 
has shown that the low-energy states of the $d$-$p$ model may be 
mapped to the $t$-$J$ ladder model with strong anisotropy between 
legs and rungs.  There exist interactions between ladders, and thus 
the generic model for the low-energy electronic states is the coupled 
anisotropic ladders described by the trellis-lattice $t$-$J$ model 
at quarter filling.\cite{nishimoto3}  This system is a Mott insulator 
because of the strong anisotropy of the ladders:\cite{nishimoto3} 
with increasing anisotropy the rung turns into an effective single 
site at half filling with a Hubbard 
repulsion.\cite{nishimoto1,nishimoto2,nishimoto3}  
However, because the anisotropy is not extremely strong, the system 
cannot be regarded as the one-dimensional (1D) Mott insulator, as is 
evident in the significant deviations of the calculated single-particle 
spectral functions for the ladders with corresponding anisotropy from 
those for the single chain,\cite{riera1,nishimoto5} although the spin 
degrees of freedom behave as a 1D Heisenberg chain with the effective 
antiferromagnetic exchange coupling.\cite{horsch,xray1,nishimoto3}  
The rungs (or dimers) with their internal hopping and exchange 
interactions are thus the essential ingredient for the basic electronic 
states of the present system.  A rung has two sites with one electron 
and is uniaxial in structure, and behaves as a `repulsive-$U$ 
center' with the effective Hubbard interaction $U_{\rm eff}=2t-J$ (for 
a $t$-$J$ dimer),\cite{nishimoto1} which however has an internal charge 
degrees of freedom; i.e., the one-electron wave function is a linear 
combination of the two tight-binding basis functions located on the 
left and right sites of the rung.  The rungs are aligned in parallel 
on the two-dimensional (2D) plane to form the trellis lattice, 
and this configuration produces a unique interplay between charge 
and spin degrees of freedom, which we believe is the source of the 
observed anomalous behaviors of NaV$_2$O$_5$.  Here the coupling 
between the rungs, especially the intersite Coulomb repulsions, plays 
an important role in the charge-ordering scenario for the mechanism of 
the phase transition.\cite{seo,nishimoto3,mostovoy}  

In this letter, we address the following question: suppose an assembly 
of such dimers and assume that the interaction between dimers is 
essentially the long-range Coulomb repulsion, then what is the charge 
dynamics of the assembly?  We take two lattice models shown in Fig.~1: 
the 2D trellis lattice consisting of anisotropic ladders as a model for 
the (V$_2$O$_3$)$^{3+}$ layer of NaV$_2$O$_5$ and the 1D lattice with 
dimerization as a model for the rungs coupled in the direction perpendicular 
to the ladders of the trellis lattice.  
We will thereby show that the intersite Coulomb repulsions between the 
dimers induce a low-energy collective mode in the charge excitations 
of the systems which is due to the internal charge degrees of freedom 
of the dimers.  We will then consider its relevance to the electronic 
states of NaV$_2$O$_5$ and present some consequences on the low-energy 
charge dynamics of this compound.  

The Hamiltonian of our model systems is given by\cite{nishimoto3} 
\begin{eqnarray}
H=&-&\sum_{\langle ij\rangle\sigma}t_{ij}
\big(\hat{c}^\dagger_{i\sigma}\hat{c}_{j\sigma}+{\rm H.c.}\big)
\nonumber \\
&+&\sum_{\langle ij\rangle}J_{ij}\Big({\mib S}_i\cdot{\mib S}_j
-\frac{1}{4}n_in_j\Big)
\nonumber \\
&+&\sum_{<ij>}V_{ij}n_in_j,
\end{eqnarray}
where 
$\hat{c}^\dagger_{i\sigma}=c^\dagger_{i\sigma}(1-n_{i,-\sigma})$ 
is the constrained electron-creation operator at site $i$ and 
spin $\sigma$ $(=\uparrow,\downarrow)$, 
$n_i=n_{i\uparrow}+n_{i\downarrow}$ is the electron-number 
operator, ${\mib S}_i$ is the spin-$\frac{1}{2}$ operator, 
and $\langle ij\rangle$ represents a pair of neighboring sites 
$i$ and $j$.  We restrict ourselves to the case of quarter 
filling.  We define the hopping and exchange parameters $t_{ij}$ 
and $J_{ij}$ as $t_\perp$ and $J_\perp$ for the rungs and 
$t_\parallel$ and $J_\parallel$ for the legs, respectively, and 
also as $t_{xy}$ and $J_{xy}$ for the zigzag-chain bonds where 
the intersite Coulomb repulsions $V_{xy}$ is also taken into 
account.  For the 1D chain model, we define the parameters for 
alternating bonds as $t_\perp$ and $J_\perp$ for stronger bonds 
and $t_{xy}$ and $J_{xy}$ for weaker bonds.  We include $V_{xy}$ 
only in the latter bonds.  
We use values of the ladder parameters obtained in 
ref.\cite{nishimoto3}: 
$t_\perp=0.298$, 
$t_\parallel=0.140$, 
$J_\perp=0.0487$, and 
$J_\parallel=0.0293$ in units of eV.  
The value $t_{xy}=0.05$ eV is assumed and the relation 
$J_{xy}=4t_{xy}^2/U_d$ is used for simplicity to obtain the value 
of $J_{xy}$.  The value of $V_{xy}$ is then varied for simulating 
various situations.  The repulsion $V_\parallel$ on the leg bonds 
is also taken into account for some discussions.  
We employ a Lanczos exact-diagonalization technique on small clusters 
to calculate the ground state and excitation spectra of the systems.  
We take a cluster of coupled four $2\times 2$ ladders (or $2\times 2$ 
unit cells) with periodic boundary condition for the trellis-lattice 
model and a 16-site (or 8 unit cell) ring for the 1D dimerized lattice 
model.  
\begin{figure}
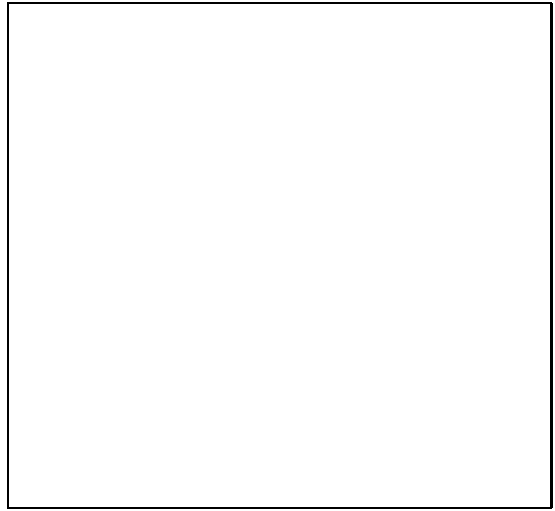

\figureheight{6.5cm}
\caption{
Schematic representation of the lattice structures of 
(a) the trellis lattice $t$-$J$ model and 
(b) the 1D dimerized $t$-$J$ model.  
The rungs in (a) and dimers in (b), which have the strongest hopping and 
exchange interactions, are indicated by bold lines.  
}
\label{fig:1}
\end{figure}

A direct way for examining the charge dynamics of the systems may 
be to calculate the dynamical density correlation function defined by 
\begin{equation}
C({\mib q},\omega)=\sum_{\nu>0} |\langle\psi_\nu 
|\rho_{\mibs q}|\psi_0\rangle|^2
\,\delta(\omega-(E_\nu-E_0)), 
\end{equation}
where $\rho_{\mibs q}$ is the Fourier transform of the electron-number 
operator, and $E_\nu$ and $|\psi_\nu\rangle$ are respectively the 
$\nu$-th eigenenergy and eigenstate of the system (with the ground state 
denoted by $\nu=0$).  
We have examined the individual excitations of electrons by the 
optical conductivity calculations,\cite{nishimoto4} whereas by the 
present $C({\mib q},\omega)$ calculations one can find possible 
low-energy collective excitations as well.  

\begin{figure}
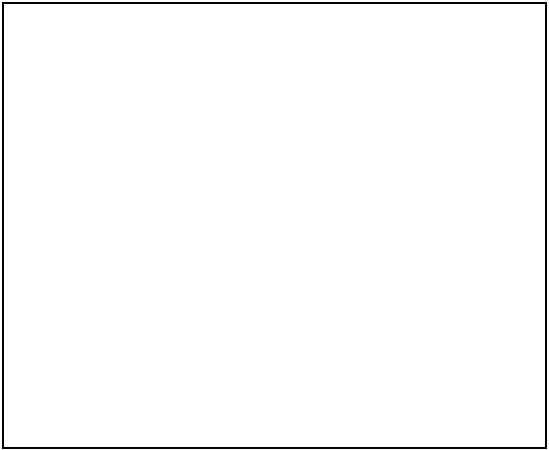

\figureheight{5.7cm}
\caption{
Dynamical density correlation function $C(q,\omega)$ calculated 
for the 1D dimerized $t$-$J$ model with the parameter values of 
$t_\perp$, $t_{xy}$, $J_\perp$, and $J_{xy}$ listed in the text.  
All the $\delta$-functions in Eq.~(2) are Lorentzian broadened with 
the width of $0.03$ eV.  
}
\label{fig:2}
\end{figure}
Now let us first examine the calculated results for the 1D dimerized 
$t$-$J$ model.  The results are shown in Fig.~2, where the parameter 
values used are in the region of strong dimerization $t_\perp/t_{xy}=5.96$ 
and thus the charge gap clearly opens.\cite{nishimoto1}  
We immediately find here that at $V_{xy}=0$ the spectra are the ones 
consistent with particle-hole transitions across the charge gap but with 
increasing the value of $V_{xy}$ there appears the low-energy peak at 
$q=0$ which is smoothly connected to the higher-energy peaks at larger 
momenta with a well-defined dispersion $\omega_{\mibs q}$, the width of 
which scales with $V_{xy}$.  We thus find a collective-mode--like behavior 
in the spectra.  Although not related to the present issue, we find 
for very large values of $V_{xy}$ ($\geqap 1$ eV) that there appears 
a set of low-energy peaks at $\omega\leqap 0.2$ eV, the position and 
weight of which are almost momentum-independent; this feature becomes 
dominant for $V_{xy}\geqap 2$ eV and is simply a consequence of the 
stabilization of the state with two electrons in a dimer, as is naturally 
expected by intuition.  

\begin{figure}
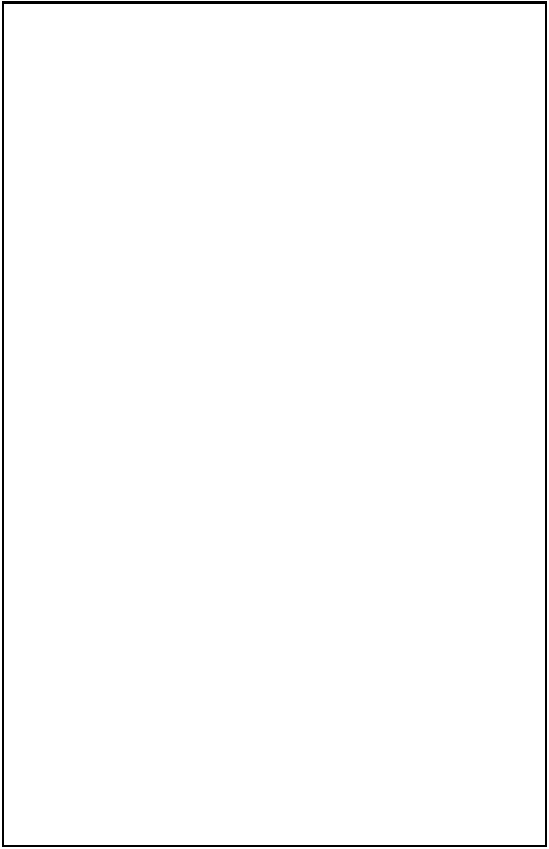

\figureheight{11.0cm}
\caption{
Dynamical density correlation function $C({\mib q},\omega)$ calculated 
for the trellis-lattice $t$-$J$ model.  
The momentum ${\mib q}=(q_\perp,q_\parallel)$ is shown in each panel.  
}
\label{fig:3}
\end{figure}
Similar spectral features are also found in the trellis lattice model 
(see Fig.~3) although the results are only for the momenta 
${\mib q}=(0,0)$, $(\pi,0)$, $(0,\pi)$, and $(\pi,\pi)$ where the spectra 
for the latter two momenta are equivalent due to symmetry.  We find that 
the features reflecting individual excitations at small $V_{xy}$ values 
gradually change into the features resembling a collective-mode excitation 
by increasing $V_{xy}$ values, i.e., the low-energy peak at ${\mib q}=(0,0)$ 
which `connects' with the higher-energy peak at ${\mib q}=(\pi,0)$ and 
$(0,\pi)$, where the width of the dispersion $\omega_{\mibs q}$ again 
scales with $V_{xy}$.  If we also include the repulsion $V_\parallel$ 
in the legs of the ladders, we find that the lowest-energy peak shifts 
to the nonzero momenta ${\mib q}=(0,\pi)$ and $(\pi,\pi)$ as shown in 
Fig.~4, the behavior being naturally expected from simple consideration 
of the repulsion between two electrons.  Thus, either `ferroelectric' 
or `antiferroelectric' fluctuation with corresponding dispersion 
$\omega_{\mibs q}$ is realized depending on the relative strength of 
the intersite repulsions; other types of fluctuations would also be 
realized if we include some longer-range repulsions.  
These observations are quite natural because the parameters we are working 
with are in the region where the long-range charge order appears in the 
infinite systems, as the calculated values of the equal-time charge 
correlations $\langle n_in_j\rangle$ directly indicate (where there 
should be a critical repulsive strength $V_c$),\cite{nishimoto3} and thus 
we should observe the features corresponding to the Goldstone mode at 
$V>V_c$, although in finite-size systems no real symmetry-breaking occurs 
and thus there is no exactly zero-energy excitation.  
\begin{figure}
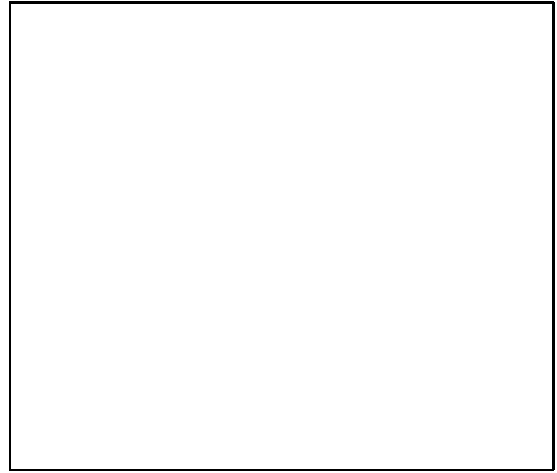

\figureheight{6.0cm}
\caption{
As in Fig.~3 but for the case with nonzero values of the repulsion 
$V_\parallel$.  
}
\label{fig:4}
\end{figure}

A speculation to what happens at $T>0$ in the thermodynamic limit 
seems to be straightforward from the results given above.  
With increasing temperature from 0 K the thermal fluctuations in the 
long-range charge-ordered state develop and at $T=T_c$ there occurs a 
phase transition.  The fluctuations of charges would be remnants of 
the quantum fluctuations obtained for small clusters at $T=0$ K where 
the charges oscillate coherently between left and right sites of the 
rungs using their internal degrees of freedom.  
The phase transition would be of a classical order-disorder type 
as in the theory of ref.\cite{thalmeier} where the result of a 
2D Ising model was used, and would thus be of the second order.  
Here we should note that the effective local-field and thus local 
order parameter for the charge ordering persists even well above 
$T_c$ as we have demonstrated via the optical conductivity 
calculations\cite{opt2,nishimoto4} where we used the term `charge 
disproportionation'.  
In other words, we have the local order parameter as the spatial 
left/right shift of the electron in the rung (which is nonvanishing 
even at high temperatures) and its thermal (and quantum) fluctuations 
characterize the electronic state of the system.   
This is thus quite unlike, e.g., the usual charge-density-wave transition 
where the normal fermi-liquid is recovered immediately above $T_c$ and 
the location of the fermi momentum ${\mib k}_{\rm F}$ plays an 
essential role.  In the present system, the large charge gap exists, 
and there is no metallic screening of long-range Coulomb interactions.  
At temperatures far above $T_c$, say at room temperature, the local 
order parameters would be in a state of rapid and random oscillations 
without coherence, but with lowering temperature, the spatial correlation 
develops, and the correlation length increases and diverges as the 
temperature approaches $T_c$ where the long range order comes 
out.  From experimental data which we have summarized 
above,\cite{ravy,nakao,fertey,opt2} one may expect that the temperature 
at which the coherent fluctuations start to develop would be in a range 
of $T\simeq 50-150$ K.  The character of the fluctuations, i.e., 
ferroelectric, antiferroelectric, or whatever, would have a consequence 
on the charge-ordering pattern at $T<T_c$, and needs to be clarified 
further from both experiment and theory.  
We should note that the lattice degrees of freedom may well play a role 
in the fluctuations of charges.  A recent study of the local Holstein 
and/or Peierls couplings in the 1D chain and ladder models\cite{riera2} 
seems quite suggestive, although nonadiabatic treatment will be 
essential for treating the slow order-parameter fluctuations with a 
time scale compatible with the lattice fluctuations.  
As for the spin degrees of freedom, it is on one hand an open 
question whether the uniform susceptibility in such a charge state 
fits well to the Bonner-Fisher curve\cite{bonner} with significant 
deviations below 250 K,\cite{hemberger} but on the other hand 
the charge-ordering pattern would naturally explain the origin of 
the spin gap at $T<T_c$\cite{seo,nishimoto3,mostovoy}.  
Actual determination of the pattern would however be an intricate 
problem because not only the long-range nature of the Coulomb 
interaction but also the lattice degrees of freedom seem to play 
an important role.\cite{mostovoy}  Suggestions from experiment are 
most desirable.\cite{chatterji} 
Finally, let us remark on the submillimeter-wave ESR experiment 
by which a direct observation of the spin gap has recently been made 
successfully.\cite{luther}  We here would like to suggest that if 
the mechanism of the spin-gap opening is due to local spin-singlet 
formation by the charge ordering\cite{ohama,seo,mostovoy} and if the 
slow fluctuations leading to the charge-ordering instability present 
well above $T_c$, the fast time-scale measurement like this ESR 
experiment could possibly detect the `spin gap' even above $T_c$ 
which is continuous to the spin gap below $T_c$.  
So far, neutron scattering experiments have not found such 
signals.\cite{fujii,yosihama}  

In summary, we have calculated the dynamical density correlation 
function of the 1D dimerized $t$-$J$ model and trellis lattice 
$t$-$J$ model at quarter filling using an exact-diagonalization 
technique on small clusters, whereby we demonstrate that the 
intersite Coulomb repulsions between the dimers (or rungs) induce 
a low-energy collective mode in the charge excitations of the systems 
where the internal charge degrees of freedom of the dimers (or rungs) 
play an essential role.  
We have argued that the electronic states and charge ordering of 
NaV$_2$O$_5$ should be understood in terms of the fluctuations 
of the local order parameters representing the electron locations 
in the rungs, and have discussed the experimental consequences 
of the fluctuations.  

We would like to thank A. N. Vasil'ev and T. Ohama for enlightening 
discussions on the experimental aspects of NaV$_2$O$_5$.  
Financial support for S.~N. by Sasakawa Scientific Research Grant 
from the Japan Science Society and for Y.~O. by Iketani Science 
and Technology Foundation are gratefully acknowledged.  
Computations were carried out at Computer Centers of the Institute 
for Solid State Physics, University of Tokyo and the Institute for 
Molecular Science, Okazaki National Research Organization.

\end{document}